# A survey of ab-initio calculations shows that segregation-induced grain boundary embrittlement is predicted by bond-breaking arguments

Michael A. Gibson and Christopher A. Schuh[*]

Department of Materials Science and Engineering, Massachusetts Institute of Technology, 77 Massachusetts Avenue, Cambridge MA 02139 USA

Contact Information:

Michael A Gibson: m_gibson@mit.edu, 617-258-5032

Christopher A Schuh: schuh@mit.edu, 617-253-6901 [*]Corresponding Author

**Abstract**

The segregation of solute atoms to grain boundaries can have a large influence on the mechanical behavior of polycrystals, particularly in metallic alloys. An overview of 400 calculations of solute-induced changes in grain boundary cohesion from 77 separate studies quantitatively demonstrates that the majority of the variation in solute-induced changes in boundary cohesion is explained by simple bond-breaking arguments. This trend is robust to changes in crystal structure and computational methods. The secondary contributions to embrittlement from other mechanisms, such as atomic size effects and charge transfer, are quantified and discussed as well.

**Keywords:** Grain Boundary Segregation; Embrittlement; Ab-Initio Calculation; Surface Segregation; Fracture

Grain boundaries (GBs) often mediate failures in engineering alloys, most commonly due to the segregation of impurities. GB embrittlement is usually associated with impurities' effects on GB cleavage, i.e. the separation of the boundary to form free surfaces. The literature contains many posited mechanisms for impurity-induced GB embrittlement, among them:

- oversize atoms may induce local tensile strains which ease boundary decohesion (i.e. atomic size effects)[1],
- low-energy bonds across the interface may permit facile surface formation [2],
- changes in the local GB electronic structure, or the formation of covalent bonds due to segregation may affect boundary cohesion [3, 4], or
- reductions in 'bond mobility' may reduce the amount of plastic deformation at the GB [5].

Certainly all of these proposed mechanisms may have relevance in specific cases, but it is interesting for alloy design purposes to develop a global, more general understanding of solute-induced changes in GB cohesion.

A seminal effort at a global understanding of GB cohesion was provided in 1980 by Seah [6], who used a regular solution-based analysis to hypothesize that solutes' effects on boundary cohesion should be determined by the intrinsic strength of the bonds formed by a solute relative to those of the base metal. Seah quantified this difference in bond strengths via the elemental sublimation enthalpies of the solute and solvent. However, at the time of Seah's work, the prediction of whether a solute would embrittle a boundary could only be validated against qualitative observations of decreases in toughness upon introduction of a given solute, and controlled studies were mostly limited to Fe-based systems. Since then, a large number of atomic-scale simulations [7-15] and more detailed experiments have been performed across a variety of systems, enabling a more quantitative, broad analysis of solutes' effects on grain boundary cohesion.

It is our purpose in this paper to first revisit the order-of-magnitude energetic analysis of Seah, in light of the large volume of quantitative data now available. We then proceed to test his bond-strength hypothesis quantitatively, and to generalize his analysis to systems beyond Fe.

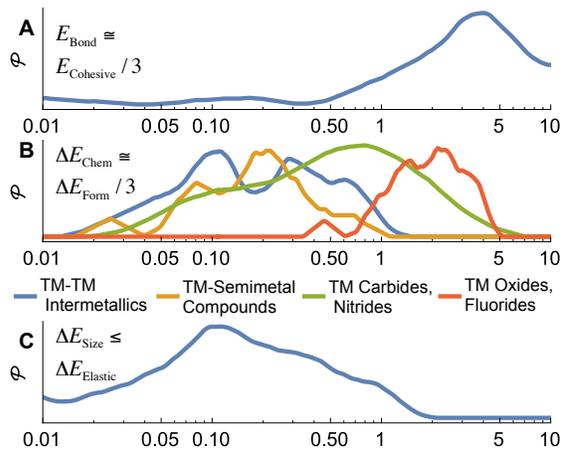

**Figure 1**: A logarithmic comparison of energy scales. **A**: The distribution, $\mathcal{P}$, of elemental cohesive energies **B**: The distribution of the energies of formation, $-\Delta E_{\text{Form}}$, of a representative set of transition metals with different classes of elements. **C**: The distribution of elastic mismatch energies for transition metal pairs.

We first reiterate and further develop Seah's original order-of-magnitude analysis. In a view not restricted to the regular solution approximation, Geng et al. posited that the change in a grain boundary's cohesive energy due to introduction of a solute, $\Delta E_{\text{B}}$, can be

approximated as a sum of effects due to bond-breaking, chemical interactions, and atomic size effects [16]:

$$\Delta E_B \cong \Delta E_{Bond} + \Delta E_{Chem} + \Delta E_{Size}. \qquad (1)$$

In the following paragraphs we estimate these three terms in turn to assess their relative importance.

First, the bond-breaking term, $\Delta E_{Bond}$, represents the change in cohesive energy simply due to the replacement of solvent-solvent bonds with solute-solute ones at the boundary upon solute segregation. It can be estimated by the difference in cohesive energies between the solute and the solvent, $\Delta E_{Bond} \cong \left(E^{coh}_{solvent} - E^{coh}_{solute}\right)/3$ [16], with the factor of three included because only a fraction of an atom's bonds are broken during creation of a surface. Fig. 1a shows the distribution of elemental cohesive energies according to Kittel [17], ranging from close to zero for the noble gases to 8.9 eV/atom for W. For typical engineering alloys, $\Delta E_{Bond}$ should usually be on the order of one to several eV/atom.

Second, the $\Delta E_{Chem}$ term represents the effects of chemical interactions on the strength of bonding across the interface; the extent to which the energy of an solute-solvent bond differs from the average of the solvent-solvent and solute-solute bond energies. Following Pauling [18] and Miedema [19], one can estimate the per-atom effects of chemical interactions between elements from the molar energies of formation of their compounds, i.e., $\Delta E_{Chem} \cong 1/3\ \Delta E_{Form}$, again introducing the factor of three to account for the fraction of bond energy lost upon decohesion. In Fig. 1b we collect data from Refs. [20, 21], and show that transition metal and transition metal-semimetal binaries typically exhibit $\Delta E_{\text{Form}}$ around 0.2 eV/atom, implying $\Delta E_{\text{Chem}} \cong 0.1$ eV/atom. (The figure also shows that chemical effects can be larger for ionic and covalent solids, such as fluorides, oxides and strong nitride formers, but for the present purposes the order of magnitude ~0.1 eV/atom is a reasonable representation for most cases).

Third, an upper bound on the atomic size effect on boundary cohesion ($\Delta E_{\text{Size}}$) can be loosely estimated by the amount of elastic energy stored due to atomic size mismatch in the crystalline bulk; this energy would be at least partially released by segregation, which is what makes this an upper bound estimate. Friedel derived the form of this energy according to continuum theory as [22]:

$$\Delta E_{\text{Size}} \leq \Delta E^{\text{elastic}} = \frac{24\pi K_A G_B r_A r_B (r_B - r_A)^2}{3 K_A r_A + 4 G_B r_B} \qquad (2)$$

Fig. 1c shows the distribution of calculated values of $\Delta E_{\text{Elastic}}$ for a representative class of transition metal solute-solvent pairs, with the values centering around 0.1 eV and reaching at most 2 eV, using data from [23] as input. While some exceptionally large elements from outside the transition metal series may exhibit larger elastic effects, e.g. the rare earths, the above analysis demonstrates that for the vast majority of solute-solvent pairs, $\Delta E_{\text{Size}}$, is expected to be lower than 0.5 eV, and for most solutes, $\Delta E_{\text{Size}}$ is of order 0.1 eV or less.

The three panels of Fig. 1 can be directly compared to the three terms of Eq. (1), and support the main hypothesis of Seah [6]: elastic and chemical effects in grain boundary embrittlement are expected to be smaller, second order effects when compared to the straightforward change in cohesive energy caused by the replacement of solvent atoms at the grain boundary. In other words, changes in boundary cohesion should be well-correlated with bond-breaking analogs.

In atomistic simulations, a solute's embrittling potency, $\Delta E_\text{B}$, is calculated as the difference in the work of separation of a grain boundary upon segregation of a solute atom, B, to the boundary:

$$\Delta E_\text{B} = W_\text{impure}^\text{sep} - W_\text{pure}^\text{sep}, \tag{3}$$

where $W_\text{impure}^\text{sep}$ and $W_\text{pure}^\text{sep}$ refer to the work of separation of the impure, segregated boundary and the pure boundary, respectively.

Using Eq. (3) and a variety of literature data, Lejček and Šob recently examined the Seah hypothesis by comparing experimental and computational embrittling potencies in Fe to elemental sublimation energies [24]. Lejček and Šob achieved an impressive correlation that is supportive of the Seah hypothesis. However, almost half of the results discussed from Ref. [24] came from an analytical model of solutes' effects on GB cohesion by Geng et al. [16], rather than a set of ab initio-computed embrittling potencies. This is problematic since, in agreement with Seah, the largest component of the analytical model of Geng et al. is the difference in the cohesive energies of the solute and solvent. As the cohesive energy is the enthalpy of sublimation in the limit as $T \to 0$, the correlation between the embrittling potencies and the difference in enthalpies of sublimation was possibly an artifact of the inclusion of the data from Geng et al.'s model in the prior study. We have found that the trend supporting the Seah hypothesis in Ref.

[7] becomes much weaker when these data are removed from consideration (i.e., the sample size-adjusted coefficient of determination, $\bar{R}^2$, reduces from 0.52 to 0.30).

In what follows, we re-create the correlation approached by Lejček and Šob in Fe-based systems by gathering additional studies, and also extend the discussion to a much broader range of metals that have been studied by ab initio methods; the bond-breaking arguments laid out above should be agnostic to the choice of base metal, but this has not, to our knowledge, been tested before.

We gathered a large set of computed embrittling potencies from the open literature and other sources, including reports to funding agencies and calculations reported in conferences from industry (a complete list of references and the data taken therefrom, as well as our data selection procedures, are collected in the online Supplemental section). The vast majority of studies were based on ab initio methods and explicitly studied solute segregation at grain boundaries in metals. In total, our survey included 77 individual studies or publications, including 400 investigations of solute sergregation for 182 unique solute-solvent pairs. The data was collected based on segregation in the dilute limit, such that the embrittling potency of the solute is considered only at the most favorable site for grain boundary segregation, but under the assumption that the solute remains at the same site during fracture (i.e. that crack propagation is fast relative to diffusion).

Prior to discussion of the data, we first emphasize: scatter must be expected when aggregating the results of so many studies, especially because they examine different grain boundaries from different crystal structures, and because the studies use different exchange-correlation functionals, different software implementations, and different simulation parameters (e.g. basis sets, pseudopotentials, and reciprocal space sampling). To avoid overplotting and to more accurately summarize trends across studies, if multiple embrittling potencies have been calculated for the same solute/solvent pair, then we take the mean of all calculations for that solute/solvent pair. Each point displayed thus represents what might be interpreted as the field's average estimate of the embrittling potency of a given solute in a given solvent.

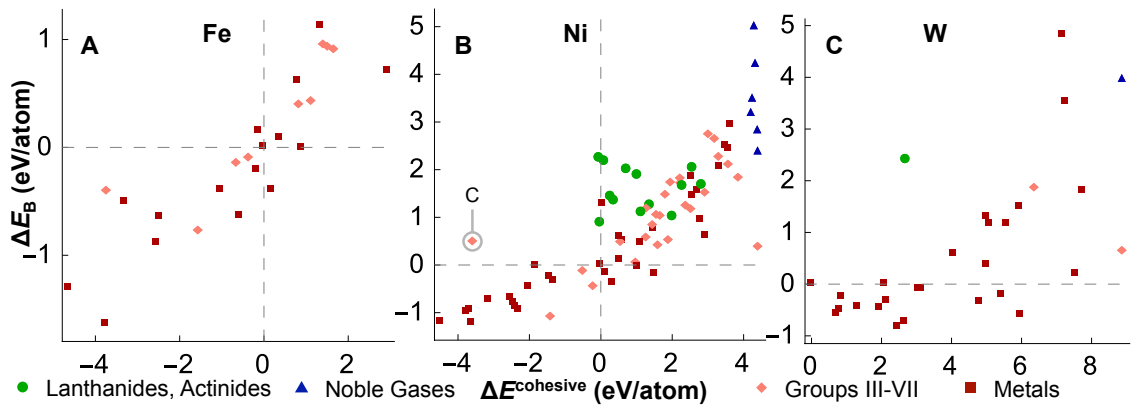

**Figure 2**: Embrittling potencies versus the difference in elemental cohesive energies for **A** Fe-, **B** Ni-, **C** and W-based systems

In Fig. 2**A** we show the embrittling potencies for solutes added to Fe GB's. A positive embrittling potency corresponds to a lower work of separation upon introduction of the impurity to the boundary, and thus to an embrittled boundary. In the spirit of the Seah hypothesis, we use as the x-axis the difference in elemental cohesive energies of the elements, $\Delta E^{\text{cohesive}} = E^{\text{cohesive}}_{\text{solvent}} - E^{\text{cohesive}}_{\text{solute}}$; data conforming to the Seah hypothesis should thus emerge as showing a strong trend on this plot. The elemental cohesive energies are used as an analogue for the difference in bond strengths as opposed to the enthalpies of sublimation proposed by Seah because segregation in ab initio studies effectively takes place at zero Kelvin.

As can be seen, there is a fairly pronounced correlation between the computed embrittling potencies and the difference in cohesive energies, demonstrating that simple bond-breaking arguments are indeed able to describe the majority of the variation in embrittling potencies for the Fe-based systems studied thus far. Further, the slope of approximately one third is consistent with the simple physical argument that approximately one third of bonds are broken during decohesion. This analysis successfully re-creates the correlation examined by Lejček and Šob [24], but benefits from reduced scatter due to our inclusion of some additional studies, our removal of the analytical data as discussed above, and our use of averaging when multiple data points were available for a single system.

In Figs. 2**B** and 2**C** we extend the analysis to solutes in Ni and W, for which the embrittling potencies show a similar trend as those for Fe, albeit over a wider range because of the larger variety of solutes studied. All of the available calculations of embrittling potencies are aggregated and presented in Fig 3**A**. The positive correlation in Fig 3**A** supports the Seah hypothesis, but there remains a significant amount of scatter. A more meaningful comparison across solvents should consider the degree to which a solute changes the cohesion of a grain boundary relative to the cohesive energy scale of the solvent, giving a dimensionless embrittling potency,

$$\Delta \varepsilon_\text{B} = \Delta E_\text{B}/\Delta E_\text{solvent}^\text{cohesive} = \left(W_\text{impure}^\text{sep} - W_\text{pure}^\text{sep}\right)/\Delta E_\text{solvent}^\text{cohesive} \quad . \tag{4a}$$

This amounts to a normalization of the y-axis in Fig 3**A**, and a similar normalization provides the form for the abscissa:

$$\Delta\varepsilon^{\text{cohesive}} = \Delta E^{\text{cohesive}}/E_{\text{solvent}}^{\text{cohesive}} \tag{4b}$$

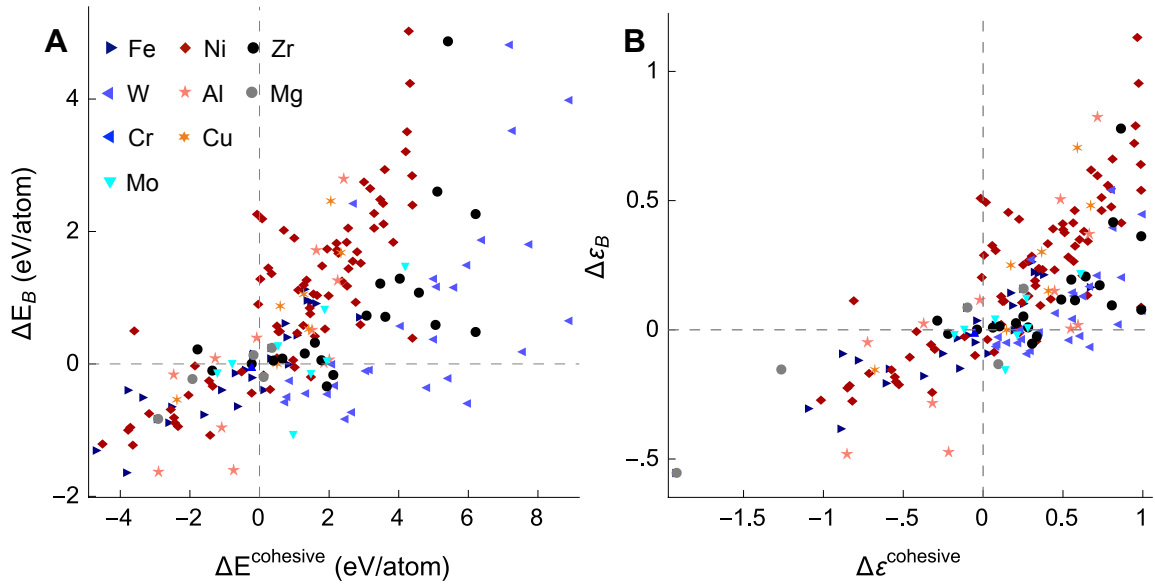

**Figure 3:** Summaries of embrittling potencies across alloy systems. **A** Embrittling potencies versus the difference in elemenental cohesive energies. **B** Dimensionless embrittling potencies of solutes at grain boundaries versus dimensionless differences in cohesive energies of the solute and the solvent. Points are colored according to the solvent.

The normalization presented in Fig. 3**B** uses the full set of data from Fig. 3**A**, and achieves a reasonable collapse of the many disparate studies into a single well-formed trend. For example, the results for Al and Mg, which were difficult to see on the absolute scale of Fig 3**A,** are now more pronounced on the normalized energy scale. Similarly, the results for Zr and W, which looked somewhat anomalous in Fig 3**A**, fall closer the trends for the other solvents once non-dimensionalized. Significantly, different solvents fall into different regions from left to right on this plot: Mg has a rather low cohesive energy, and thus many alloying elements are expected to increase GB cohesion in Mg. In contrast, W is the most cohesive element, and so it would be fairly unlikely for an element to significantly increase the cohesion of W GB's.

Non-dimensionalizing the embrittling potencies renders them on the same scale, such that it becomes reasonable to use linear regression to summarize the data. Formally, the assumptions of independent observations and a constant variance inherent to statistical interpretation of linear regression are not satisfied in the present dataset, and so this analysis merely provides a one-number summary of the variation of the data, and an estimate of the slope. We find that the dimensionless difference in cohesive energies between the solute and the solvent is able to describe 59% ($\bar{R}^2$) of the variation in dimensionless embrittling potencies via a one-parameter fit:

$$\Delta \varepsilon_B = \beta \left( \Delta E_{solvent}^{cohesive} - \Delta E_{solute}^{cohesive} \right) / \Delta E_{solvent}^{cohesive} \tag{6}$$

with $\beta = 0.41 \pm .05$. This slope is somewhat higher than the 1/3 expected from simple physical arguments described earlier. This is possibly due to the covariance of other chemical and/or elastic effects with $\Delta \varepsilon_B$, which we discuss in greater detail below.

As we noted at the outset, aggregation of data from so many sources should be expected to lead to significant scatter due to, e.g., procedural differences. However, some of the deviations from the trend in Figs. 2 and 3 likely reflect true physics: the many systems presented exhibit different chemical and elastic effects on decohesion. Following Fig. 1, such effects, while not dominant, should play a smaller but non-negligible role. Evidence for this as a source of scatter can be seen by comparing the various noble gas atoms as solutes in Ni (the right side of Fig 2**B**), which should behave identically in terms of their chemical interactions with Ni, and also exhibit similar cohesive energies near zero. The variation in their embrittling potencies (which span a factor of two) should thus be mainly due to the scatter intrinsic to the data and the different atomic sizes of the noble gases.

We can gain additional understanding of such second-order effects by examining those elements that fall well above the general trend. For example, the rare-earths, in the middle of Fig 2**B**, and carbon, on the far left. For the rare-earth elements this is almost certainly due to their oversize radii relative to that of Ni leading to large amounts of elastically stored energy ($r_{Ni}/r_{RE} = 1.39$ to $1.55$) after segregation to the GB. For C, the positive deviation is likely due to the anomalously negative cohesive energy of graphite as well as possible non-linearities due to unfavorable chemical interactions between Ni and C; Ni and C do not form any carbides at equilibrium.

In summary, we find that the bond-breaking arguments made by Seah [6] 35 years ago for Fe-based systems are a general means to summarize the hundreds of ab initio computed changes in grain boundary cohesion due to solute segregation across alloy systems. Such data were not available to Seah, and previous studies have mostly examined small subsets of observations within a single base metal, such that global trends could not be fully revealed. Our aggregation of the data produced by the field to this point shows that when the data is considered in its totality and properly normalized, it convincingly conforms to the Seah analysis. What is more, these bond breaking trends are apparently robust to variations in crystal structure and computational methodology, with the majority of the variation in embrittling potencies explained based solely on bond-breaking arguments. Deviations from these trends are due to a combination of chemical and atomic size effects, both of which present significant but generally small effects on embrittlement. Our findings, as well as the database presented in the online supplement, may be useful in alloy design across metallic systems.

**Acknowledgments**

This work was supported by the US Army Research Office, under grant W911NF-14-1-0539, and through the Institute for Soldier Nanotechnologies at MIT. M.A.G. was supported by the Department of Defense (DoD) through the National Defense Science & Engineering Graduate Fellowship (NDSEG) Program. We thank J. Sebastian at Questek Innovations LLC for providing data, and A. Kalidindi and A. Lai, both of MIT, for thoughtful discussions.

**Captions**

**Figure 1**: A logarithmic comparison of energy scales. **A**: The distribution, $\mathcal{P}$, of elemental cohesive energies **B**: The distribution of the energies of formation ,$-\Delta E_{\text{Form}}$, of a representative set of transition metals with different classes of elements. **C**: The distribution of elastic mismatch energies for transition metal pairs.

**Figure 2**: Embrittling potencies versus the difference in elemental cohesive energies for **A** Fe-, **B** Ni-, **C** and W-based systems

**Figure 3:** Summaries of embrittling potencies across alloy systems. **A** Embrittling potencies versus the difference in elemenental cohesive energies. **B** Dimensionless embrittling potencies of solutes at grain boundaries versus dimensionless differences in cohesive energies of the solute and the solvent. Points are colored according to the solvent.

# Figure 1

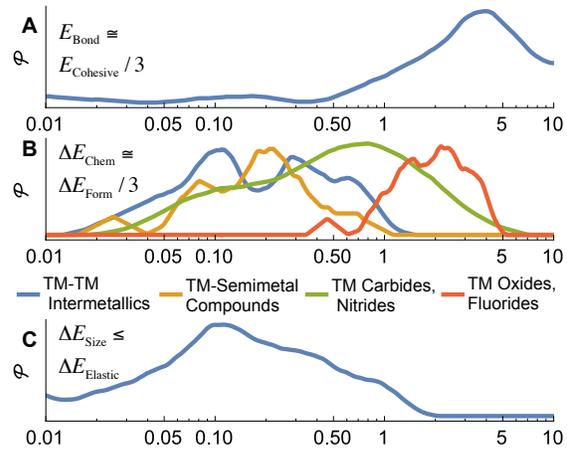

**Figure 2**

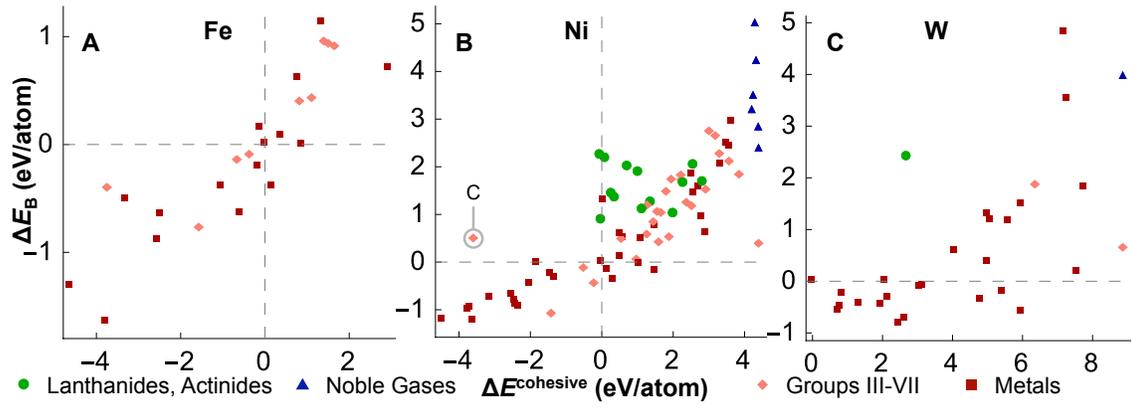

**Figure 3**

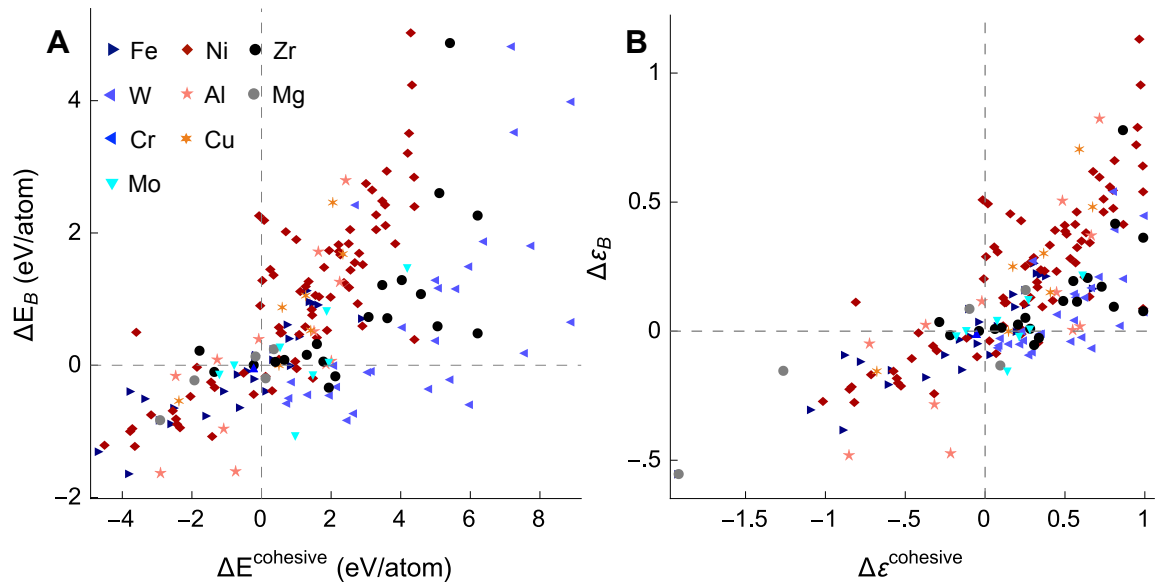